\documentstyle[psfig,conf_iap,]{article}
\newcommand{\lya}{Ly$\alpha$}

\newcommand{\HI}{{H\sc{i}}}
\newcommand{\civ}{{C\sc{iv}}}
\newcommand{\nh}{N_{\rm HI}}

\newcommand\cdunits{{\rm cm}^{-2}}
\newcommand\ggh{{\Gamma_{\rm HI} }}

\begin{document}
\heading{ The Metallicity of the Local IGM from the HST/STIS
Spectrum of 3C273}
\par\medskip\noindent
\author{%
Romeel Dav\'e$^{1}$, S. R. Heap$^{2}$, G. M. Williger$^3$, R. J. Weymann$^4$, T. M. Tripp$^5$, E. B. Jenkins$^5$
}
\address{ Steward Observatory, 933 N. Cherry Ave., Tucson, AZ, 85721 }
\address{ Goddard Space Flight Center, Greenbelt, MD, 20771}
\address{ Astronomy Dept., John Hopkins Univ., 3400 N. Charles St., Baltimore, MD, 21218}
\address{ Carnegie Observatories, 813 Santa Barbara St., Pasadena, CA, 91101}
\address{ Princeton University Observatory, Princeton, NJ, 08544}

\begin{abstract}
We present a preliminary study of the metallicity of low-redshift
Lyman alpha absorbers from the HST/STIS spectrum of 3C273.  Using a
pixel-based shift-and-coadd technique, we compare observations to
carefully-constructed mock quasar spectra from a cosmological hydrodynamic
simulation.  We place an upper limit of [C/H]$_\odot < -1.5$ at $>90$\%
confidence using two \lya\ absorbers with $\nh>10^{14}\cdunits$, from
the fact that we see little absorption near the expected \civ\ positions,
whereas the artificial spectra predict significant absorption.  We assume
a Haardt \& Madau (quasar-based) ionizing background spectrum, but a
softer spectrum results in a stronger constraint on the metallicity.
With only two absorbers and various modeling uncertainties, it is too
early to make a definitive determination of the diffuse IGM metallicity,
but upcoming observations should improve constraints.
\end{abstract}
\section{Introduction}

Weak Lyman alpha (\lya) absorbers that are now detectable
in HST/STIS quasar spectra are predicted to arise in diffuse
non-equilibrium large-scale structures, analogous to high-redshift
forest absorbers~\cite{dav99}; this has been preliminarily confirmed by
first comparisons of the statistical properties of \lya\ absorbers with
simulations~\cite{dav01}.  At high redshift, absorbers with $\nh\approx
10^{14.5}\cdunits$ are already enriched to a level of [C/H]$_\odot \approx
-2.5$~\cite{son96}\cite{dav98}.  It is of great interest to study similar
absorbers at low redshift, to determine the rate at which metals have
been injected into the IGM since $z\sim 3$.  The results have implications
for understanding early galaxy formation, associated blowout, and winds.

To study absorbers equivalent to $\nh\approx 10^{14.5}\cdunits$ at $z=3$
requires one to examine absorbers with $\nh\approx 10^{13}\cdunits$
at $z\approx 0$~\cite{dav99}.  Unfortunately, since the sensitivity
of HST/STIS spectra are now only approaching Keck/HIRES and VLT/UVES,
it will be very difficult to detect metals in these systems unless the
present day metallicity is much higher.  This appears unlikely, since
the metallicity measured from archival HST/FOS spectra is [C/H]$_\odot =
-1.6$ for significantly stronger absorbers~\cite{bar98} (equivalent width
$W_r=0.45-0.75$\AA, corresponding very roughly to $\nh>10^{15}\cdunits$),
and simulations predict that less dense gas (i.e.  weaker absorbers) has
lower metallicity~\cite{cen99}.  A mitigating factor is that overdensities
of $\sim 10$ at $z=3$ correspond to overdensities of $\sim 50$ at $z=0$
due to gravitational growth of perturbations.

So instead, here we use stronger absorbers ($\nh> 10^{14}\cdunits$)
to make a first attempt at constraining the local IGM metallicity from
the HST/STIS spectrum of 3C273.  The metallicity in these systems is
expected to be [C/H]$_\odot>-2.5$, since they are have
overdensities $\approx 30-50$~\cite{dav99}, and have had significant time
since $z\sim 3$ to be enriched further.  We use \civ\ absorption to
trace the metallicity, as it is expected to be the strongest metal line
redwards of \lya\ arising from diffuse IGM gas.

\section{Data and Simulations}

The observations of 3C273 are discussed by Heap et al. in these
proceedings.  They were taken using HST/STIS's E140M grating, yielding
3 km/s pixels with 2.2 pixels per resolution element.  The average S/N
was 13 per pixel in the \civ\ region for 3C273, with a \civ\ redshift
coverage of $\Delta z=0.097$.

We compare to mock 3C273 quasar spectra drawn from a cosmological
hydrodynamic simulation~\cite{dav01}.  To model metal absorption, we
add a uniform metallicity to the gas, and compute the \civ\ absorption
using CLOUDY, taking the local density and temperature of gas from the
simulation~\cite{dav98} and assuming an ionizing background predicted from
quasars~\cite{haa96} (hereafter HM), with its amplitude set by matching
the \HI\ column density distribution amplitude~\cite{dav01}.

3C273 has some fortunate advantages for the study of metal absorption.
In addition to being at a redshift ($z_{\rm em}=0.155$) that provides
good coverage for both \lya\ and \civ, and being the brightest nearby
quasar, it also has two (apparently intergalactic) \lya\ absorbers wih
$\nh>10^{14}\cdunits$, where we would expect less than one~\cite{dav01}.
There is a strong Virgo absorber ($\nh=10^{15.8}\cdunits$) in this
spectrum that we do not consider, because it shows many low ions that
suggest it is not arising in intergalactic gas.

\section{Results}

\begin{figure}
\centerline{\vbox{
\psfig{figure=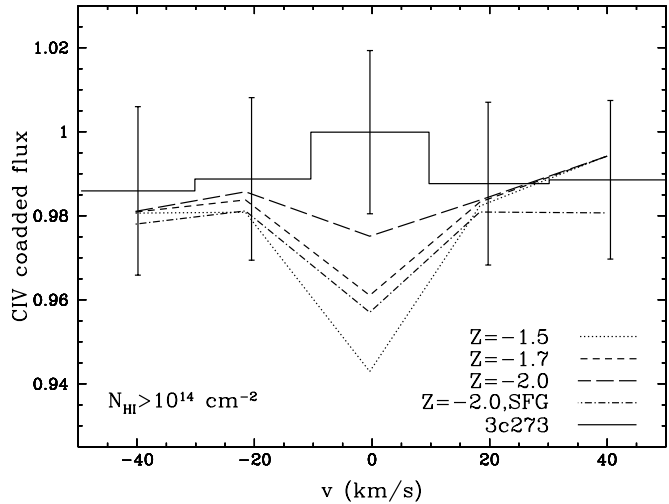,height=7.cm,angle=270}
}}
\caption[]{Pixel-by-pixel coaddition within $\pm 40$~km/s centered
on \lya\ absorbers with $\nh>10^{14}\cdunits$.  Histogram shows
result from 3C273, with other lines show results from artificial 
spectra with uniform metallicity added as indicated.
}
\end{figure}

There are no intergalactic \civ\ absorbers detected as individual lines
in the spectrum of 3C273.  Thus we must rely on stacking techniques to
improve our sensitivity.  It has been noted that stacking suffers from
the caveat that \civ\ and \lya\ absorbers can be offset by $\sim 20$~km/s
or more, so a coadded \civ\ ``feature" will appear broadened~\cite{ell00}.
This can result in an inaccurate estimate of the \civ/\HI\ ratio if not
interpreted correctly.  The reason for this offset is that the ionization
conditions of \civ\ and \HI\ do not exactly match, so when one probes
through a filament, the maximum \HI\ absorption will not necessarily
be coincident with that of \civ.  The simulations used here reproduce
this effect, since the ionization structure is computed locally across
each absorbing structure.  Thus it is important to use simulations to
interpret stacked spectra.

Our procedure is as follows: We identify each \lya\ absorber
with $\nh>10^{14}\cdunits$; 3C273 has two such \lya\ absorbers
at $\lambda=1219.77, 1296.57$\AA, having column densities of
$10^{14.2}, 10^{14.07}\cdunits$.  Our AutoVP-fit
column densities for these absorbers are consistent with results
from a FUSE+GHRS analysis~\cite{sem01}, which uses Ly$\beta$ absorption to
constrain the column density.  We examine the pixels at associated \civ
(1548.2\AA) within $\pm 40$~km/s of the \lya\ absorber velocity, and bin
those flux decrements (constrained by twice the doublet position's flux
decrement) in velocity space.

The solid line in Figure~1 shows the result of this procedure for 3C273,
while other lines show the result of varying the metallicity in the
artificial spectra from [C/H]$_\odot=-1.5\rightarrow -2.0$.  The two
absorbers in 3C273 are consistent with no absorption at the expected
\civ\ position.  [C/H]$_\odot=-1.5$ is ruled out at $>2\sigma$ level.
Also, a K-S test of the (pre-binned) flux distribution finds a 0.2\%
probability that the simulated and observed fluxes are drawn from the
same underlying distribution.  The observed and simulated samples are
formally consistent for [C/H]$_\odot=-2.0$, and discrepant by $\approx
1.5\sigma$ for [C/H]$_\odot=-1.7$.  Thus we quote a $2\sigma$ upper
limit of [C/H]$_\odot<-1.5$ for these two absorbers.

The shape and amplitude of the ionizing flux is a large uncertainty
in our modeling.  The amplitude is reasonably well-constrained at
the Lyman limit~\cite{dav01}, but altering the shape can change the
\civ\ photoionization rate significantly.  We have assumed a $J_\nu$
shape taken from quasars by HM, but star forming galaxies may provide
a significant contribution to the local flux~\cite{shu99}.  In order to
test this, we generate a ``toy" star forming galaxy (SFG) spectrum, where
we simply multiply the HM spectrum by $\nu^{-2}$, keeping $\ggh$ fixed.
The result for this ionizing spectrum is shown as the dot-dashed line in
Figure~1 for [C/H]$_\odot=-2.0$.  Qualitatively, softening the spectrum
has the effect of increasing \civ\ absorption and thereby strengthening
our metallicity constraint.

While there are only two absorber in this spectrum with
$\nh>10^{14}\cdunits$, we could attempt to lower the column density
threshold to include more lines and improve statistics.  However, even
reducing to $\nh=10^{13.5}\cdunits$ significantly weakens the result,
because the simulations predict many more pixels with little or no \civ\
absorption.  Also, note that if there was chance coincident absorption
near the \civ\ position, it would only strengthen our constraint.

\section{Summary and Discussion}

We have placed a preliminary upper limit on the metallicity in
filamentary large-scale structure (overdensity $\approx 30-50$) using
two absorbers with $\nh>10^{14}\cdunits$ in the HST/STIS spectrum of
3C273.  Using cosmological hydrodynamic simulations to calibrate the
pixel-based stacking technique, we find [C/H]$_\odot<-1.5$ at greater
than 90\% confidence level, with some sensitivity to the assumed ionizing
background shape.  Or result could be compromised if (i) our simulations
are incorrectly modeling the density-temperature structure of the IGM,
(ii) the ionizing background is significantly {\it harder} than that from
quasars, or (iii) these two absorbers have anomalously low metallicities.
While none seem {\it a priori} likely, more data and model comparisons
are needed.

If our result holds up with upcoming larger and more sensitive samples, it
would suggest that the metallicity of the diffuse IGM has increased by
less a factor of 10 since $z\sim 3$ (90\% of the age of the Universe).
This favors a scenario where the enrichment of the IGM occured
predominantly at $z\gg 3$, a fact that may constrain wind and ejection
models from high redshift galaxies (e.g. Ferrara, these proceedings),
and help interpret O{\sc{vi}} observations in the local IGM (Tripp et al.,
these proceedings).

\vfill
\end{document}